\documentclass[reprint,amsmath,amssymb,aps,superscriptaddress]{revtex4-2}
\usepackage{graphicx}
\usepackage{dcolumn}
\usepackage{bm}
\usepackage{amsmath}
\usepackage[T1]{fontenc}
\usepackage{xcolor}
\usepackage{ulem}

\begin{document}

\title{Observation of Altermagnetic Order Switching in Bulk MnTe by Polarized Neutron Diffraction}
\author{Zheyuan Liu}
\affiliation{Institute for Solid State Physics, the University of Tokyo, Kashiwa 277-8581, Japan}
\author{Shinichiro Asai}
\affiliation{Institute for Solid State Physics, the University of Tokyo, Kashiwa 277-8581, Japan}
\author{Shingo Takahashi}
\affiliation{Institute for Solid State Physics, the University of Tokyo, Kashiwa 277-8581, Japan}
\author{Hiraku Saito}
\affiliation{Institute for Solid State Physics, the University of Tokyo, Kashiwa 277-8581, Japan}
\author{Taro Nakajima}
\affiliation{Institute for Solid State Physics, the University of Tokyo, Kashiwa 277-8581, Japan}
\affiliation{RIKEN Center for Emergent Matter Science (CEMS), Saitama 351-0198, Japan}
\affiliation{Institute of Materials Structure Science, High Energy Accelerator Research Organization, Ibaraki 305-0801, Japan}
\author{Takatsugu Masuda}
\affiliation{Institute for Solid State Physics, the University of Tokyo, Kashiwa 277-8581, Japan}
\affiliation{Institute of Materials Structure Science, High Energy Accelerator Research Organization, Ibaraki 305-0801, Japan}
\affiliation{Trans-scale Quantum Science Institute, the University of Tokyo, Tokyo 113-0033, Japan}

\date{\today}

\begin{abstract}
Altermagnetic order, characterized by the N\'{e}el vector, breaks time-reversal symmetry (TRS) even in the nonrelativistic limit. Although spin-polarized and anomalous transport phenomena emerge with this order, they are mutually compensated by TRS-connected antiphase domains with opposite N\'{e}el vectors. Here we employ polarized neutron diffraction to directly probe the altermagnetic order in MnTe. Pronounced nuclear-magnetic interference terms were observed, providing direct evidence of a net N\'{e}el vector in the bulk crystal. Moreover, a weak ferromagnetic moment (WFM), originating from relativistic spin-orbit coupling, was found to be coupled with the altermagnetic order. Both the altermagnetic order and the WFM can be switched by milli-Tesla-scale magnetic field cooling.
\end{abstract}

\maketitle

% \textit{Introduction}---
Breaking time-reversal symmetry (TRS) in magnetic materials enables efficient spin-charge conversion~\cite{uchidaObservationSpinSeebeck2008,tserkovnyakEnhancedGilbertDamping2002,valenzuelaDirectElectronicMeasurement2006,kimataMagneticMagneticInverse2019,boseTiltedSpinCurrent2022,gonzalez-hernandezEfficientElectricalSpin2021}, underpinning their spintronic functionalities. Recent classification based on spin-group symmetry identified a new class of TRS-breaking magnets~\cite{smejkalConventionalFerromagnetismAntiferromagnetism2022,yuanPredictionLowZCollinear2021,hayamiMomentumDependentSpinSplitting2019}, dubbed altermagnets, which are promising material for generating giant spin current under stray-field-free condition. 
A range of exotic phenomena has been experimentally verified in altermagnets, including spin-split electronic bands~\cite{krempaskyAltermagneticLiftingKramers2024,leeBrokenKramersDegeneracy2024,osumiObservationGiantBand2024,dingLargeBandSplitting2024,chilcoteStoichiometryInducedFerromagnetismAltermagnetic2024,jiangMetallicRoomtemperatureDwave2025,zhangCrystalsymmetrypairedSpinValley2025}, chiral-split magnon band~\cite{liuChiralSplitMagnon2024,sunObservationChiralMagnon2025}, anomalous Hall effect (AHE)~\cite{gonzalezbetancourtSpontaneousAnomalousHall2023,kluczykCoexistenceAnomalousHall2024,chilcoteStoichiometryInducedFerromagnetismAltermagnetic2024,reichlovaObservationSpontaneousAnomalous2024,zhouManipulationAltermagneticOrder2025,takagiSpontaneousHallEffect2025}, and other experimental signatures of TRS-breaking~\cite{harikiXRayMagneticCircular2024,aoyamaPiezomagneticPropertiesAltermagnetic2024}. These properties inherently rely on altermagnetic order, thus only emerge when the distribution of altermagnetic antiphase domains is imbalanced in real materials. In the absence of net magnetization, whether the altermagnetic order can be directly controlled by external magnetic fields remains a crucial challenge on the pathway to functional altermagnetic devices.

The minimal model of altermagnetism consists of antiferromagnetically arranged electron spins embedded in a nonmagnetic atomic or orbital environment with rotational symmetry~\cite{roigMinimalModelsAltermagnetism2024,leebSpontaneousFormationAltermagnetism2024}. 
In the presence of additional effects such as spin--orbit coupling (SOC)~\cite{mcclartyLandauTheoryAltermagnetism2024,mazinOriginGossamerFerromagnetism2024,autieriStaggeredDzyaloshinskiiMoriyaInteraction2025,roigQuasisymmetryConstrainedSpinFerromagnetism2025,joWeakFerromagnetismAltermagnets2025}, lattice defect or distortion~\cite{alperinPolarizedNeutronStudy1962,felcherAntiferromagneticDomainsSpinflop1996,friesRemanentMagnetizationDilute1993}, and strain~\cite{disaPolarizingAntiferromagnetOptical2020,maMultifunctionalAntiferromagneticMaterials2021,aoyamaPiezomagneticPropertiesAltermagnetic2024,komuroRevisitingPiezomagneticEffect2025}, the degeneracy between altermagnetic antiphase domain states can be lifted by external magnetic fields in contrast with conventional antiferromagnets. 
SOC activates staggered Dzyaloshinskii-Moriya (DM) interactions in altermagnets, giving rise to a weak ferromagnetic moment (WFM) coupled to the altermagnetic order by spin canting~\cite{moriyaAnisotropicSuperexchangeInteraction1960,chilcoteStoichiometryInducedFerromagnetismAltermagnetic2024,kluczykCoexistenceAnomalousHall2024,autieriStaggeredDzyaloshinskiiMoriyaInteraction2025}. Anisotropic $g$-tensors with rotational symmetries lead to the WFM %in rutile and Mn₃Sn-type altermagnets 
originating from orbital magnetization~\cite{sandratskiiRoleOrbitalPolarization1996,smejkalCrystalTimereversalSymmetry2020,joWeakFerromagnetismAltermagnets2025}. Early studies showed that the altermagnetic antiphase domains in MnF$_2$ were selected by field cooling~\cite{alperinPolarizedNeutronStudy1962,felcherAntiferromagneticDomainsSpinflop1996}, likely due to lattice defects. 
Moreover, chemical doping induces WFM in MnF$_2$ via lattice distortion~\cite{friesRemanentMagnetizationDilute1993}. 
Piezomagnetic effect in altermagnets induces or enhances WFM as well~\cite{disaPolarizingAntiferromagnetOptical2020,maMultifunctionalAntiferromagneticMaterials2021,aoyamaPiezomagneticPropertiesAltermagnetic2024,komuroRevisitingPiezomagneticEffect2025}, suggesting the possible domain selection by combined application of external strain and magnetic fields.

\begin{figure*}[htbp]
	\includegraphics[width=\linewidth]{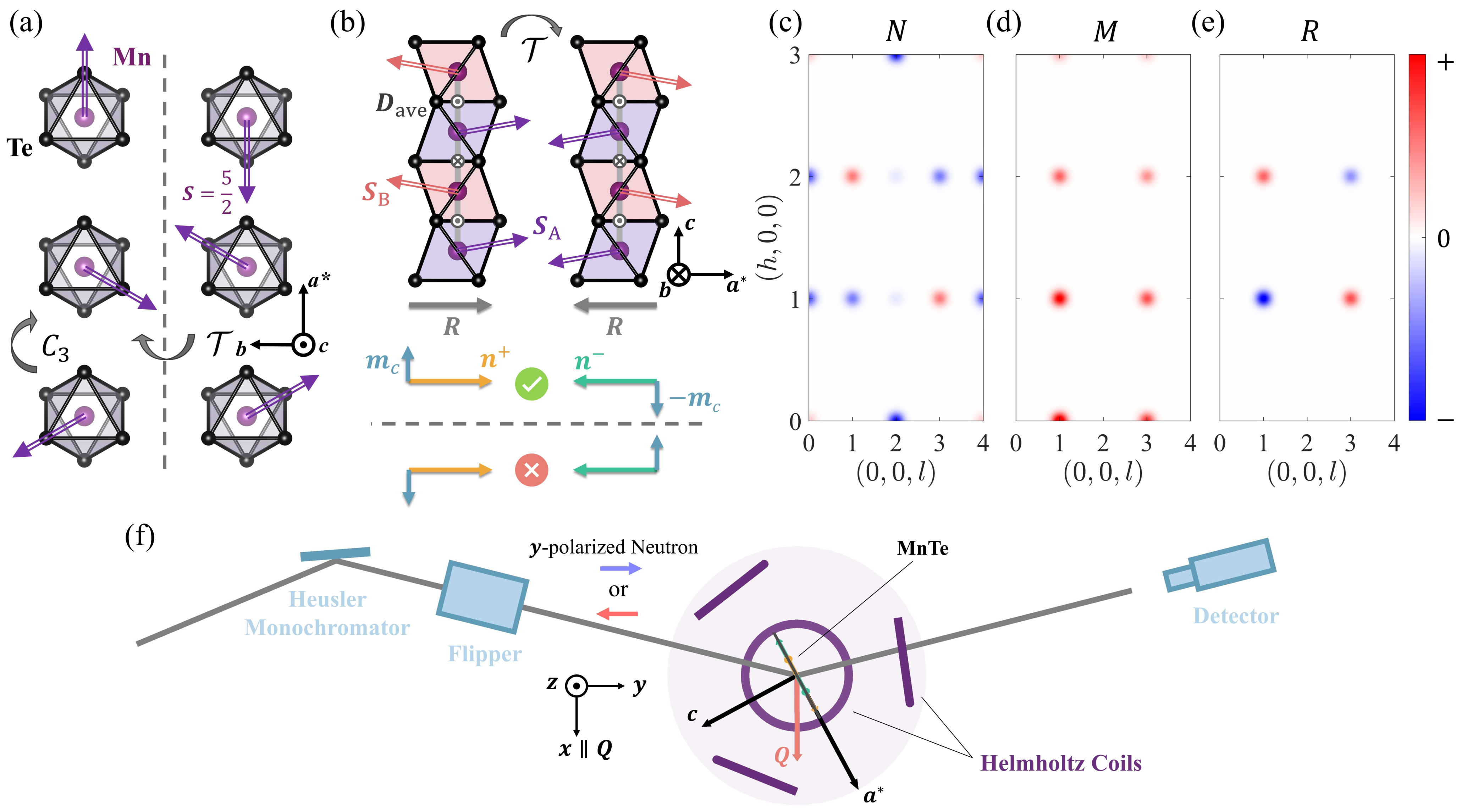}
	\caption{\label{f1}
		(a) Six altermagnetic domain states schematically shown for MnTe. The columns show orientational domain states related by a $C_3$ rotation, and the rows show antiphase domain states related by time reversal $\mathcal{T}$.
		(b) Two antiphase domain states (upper panel) with N\'{e}el vectors $\bm{n^+}$ and $\bm{n^-}$ parallel to the $a^*$-axis. The NMI vector $\bm{R}$ aligns with the N\'{e}el vector and thus changes sign between the antiphase domain states. Spins are canted due to the alternating averaged DM vector $\bm{D}_\text{ave}$, leading to a finite net moment $\bm{m}_c$. The canting angle is exaggerated for illustration. The spin configuration in the lower panel is forbidden.
		Calculated structure factors of (c) nuclei, (d) spins, and (e) the scalar part of the NMI vector in the $(h0l)$ plane.
		(f) Schematic of the PoND experimental setup in the half-polarized mode. The incident beam is polarized, while the final beam is not analyzed. A pair of vertical and three horizontal Helmholtz coils are installed at the sample position to generate the guide field as well as the cooling field along arbitrary directions.}
\end{figure*}

Antiphase domain states in conventional collinear antiferromagnets are identical in the reciprocal space when SOC is absent, and are thus only detectable in the real space. 
By contrast, due to TRS breaking, antiphase domain states in altermagnets are distinguishable even in reciprocal space~\cite{mcclartyObservingAltermagnetismUsing2025,alperinPolarizedNeutronStudy1962,felcherAntiferromagneticDomainsSpinflop1996,nathansPolarizedNeutronStudyHematite1964,almeidaMagnetisationDensityCovalency1988,kimImagingAntiferromagneticAntiphase2018}. For neutrons scattered by both the ordered electron spins and nuclear ligands in altermagnets, the observable manifests itself as nuclear--magnetic interference (NMI) terms, which alter their sign between antiphase domain states. The sign change reflects the distinguishability of time-reversal-related antiphase domains in reciprocal space, which is a defining property of altermagnets. In contrast, in conventional collinear antiferromagnets where opposite-spin sublattices are connected by time reversal combined with translation or inversion, such NMI signals are not expected, as rigorously demonstrated in Ref.~\cite{mcclartyObservingAltermagnetismUsing2025}. The NMI effect was conventionally utilized for an accurate measurement of the magnetic moment distribution in ferromagnets~\cite{mookMagneticMomentDistribution1966} because it amplifies the detection on faint magnetic moments down to the 10$^{-3} \mu_\text{B}/$atom scale~\cite{prokesPolarizedNeutronDiffraction2019}. 
In early polarized neutron diffraction (PoND) experiments, the NMI term was observed in magnets which are currently classified as altermagnets~\cite{alperinPolarizedNeutronStudy1962,felcherAntiferromagneticDomainsSpinflop1996,nathansPolarizedNeutronStudyHematite1964,almeidaMagnetisationDensityCovalency1988}, whose magnitude is comparable to that of nuclear and magnetic diffraction. 
Here, by performing PoND on the altermagnetic prototype MnTe, we observed pronounced NMI terms, representing a direct probe of the altermagnetic order in altermagnet. 
The switching of altermagnetic order, along with the WFM in MnTe, was realized by milli-Tesla--scale magnetic-field cooling.

% \textit{Weak Ferromagnetism and Domain Selection}---
MnTe has been confirmed as a prototype altermagnet by photoemission spectroscopy~\cite{krempaskyAltermagneticLiftingKramers2024,leeBrokenKramersDegeneracy2024,osumiObservationGiantBand2024} and inelastic neutron scattering~\cite{liuChiralSplitMagnon2024} in recent studies. MnTe crystallizes in a centrosymmetric hexagonal lattice with space group $P6_3/mmc$~\cite{efremdsaLowtemperatureNeutronDiffraction2005}, and no change in the lattice symmetry has been reported across the N\'eel temperature $T_{N}$~\cite{liuStraintunableAnomalousHall2025}. Below $T_N$, an $A$-type antiferromagnetic ground state with spins oriented along the <$1\bar{1}0$> direction is realized in MnTe~\cite{harikiXRayMagneticCircular2024,aoyamaPiezomagneticPropertiesAltermagnetic2024}. Two types of magnetic domain states coexist: the orientational domain states related by the $C_\text{3}$ lattice symmetry and the antiphase domain states connected by TRS, as shown in Fig.~\ref{f1}(a). 
In the centrosymmetric lattice of MnTe, the local DM interactions are compensated in the unit cell. 
However, in the presence of SOC, the feedback from altermagnetic order on the charge distribution lowers the lattice symmetry, allowing an uncompensated DM interaction which connects the adjacent Mn layers~\cite{mazinOriginGossamerFerromagnetism2024}. This DM interaction can be understood, at the level of general symmetry considerations, as an effective four-spin interaction~\cite{brinkerChiralBiquadraticPair2019,laszloffyMagneticStructureMonatomic2019}. The averaged interlayer DM vector, $\bm{D}_\text{ave}$, lies perpendicular to the N\'{e}el vector, $\bm{n}=\bm{S}_\text{A}-\bm{S}_\text{B}$, and alternates along the $c$-axis, resulting in a canted moment, $\bm{m}_c=\bm{S}_\text{A}+\bm{S}_\text{B}$, aligned with the $c$-axis (see Section V in Supplemental Material (SM)~\cite{SM} for details). 
The vector $\bm{D}_\text{ave}$ ensures that the opposite $\bm{m}_c$ corresponds to the opposite $\bm{n}$, as shown in the upper panel of Fig.~\ref{f1}(b), and forbids the spin configuration in the lower panel. A route for achieving a single magnetic domain in MnTe is to utilize the spin-flop transition~\cite{kriegnerMultiplestableAnisotropicMagnetoresistance2016} and the flipping of WFM. The WFM in MnTe is around 10$^{-5} \mu_\text{B}/$Mn along the $c$-axis~\cite{kluczykCoexistenceAnomalousHall2024}, producing almost negligible stray fields (10$^{-3}$ mT, typically 10 mT in ferromagnetic devices), yet enabling the selection of antiphase domain states by external fields. In addition, antiphase domain states in the thin film MnTe were shown to be selected by field cooling along the $c$-axis~\cite{aminNanoscaleImagingControl2024,harikiXRayMagneticCircular2024}. 
% The hysteresis of AHE in MnTe suggests the selection of antiphase domain states by applying a magnetic field along the $c$-axis as well~\cite{gonzalezbetancourtSpontaneousAnomalousHall2023,kluczykCoexistenceAnomalousHall2024,chilcoteStoichiometryInducedFerromagnetismAltermagnetic2024}. 

\begin{figure*}[htbp]
	\includegraphics[width=\linewidth]{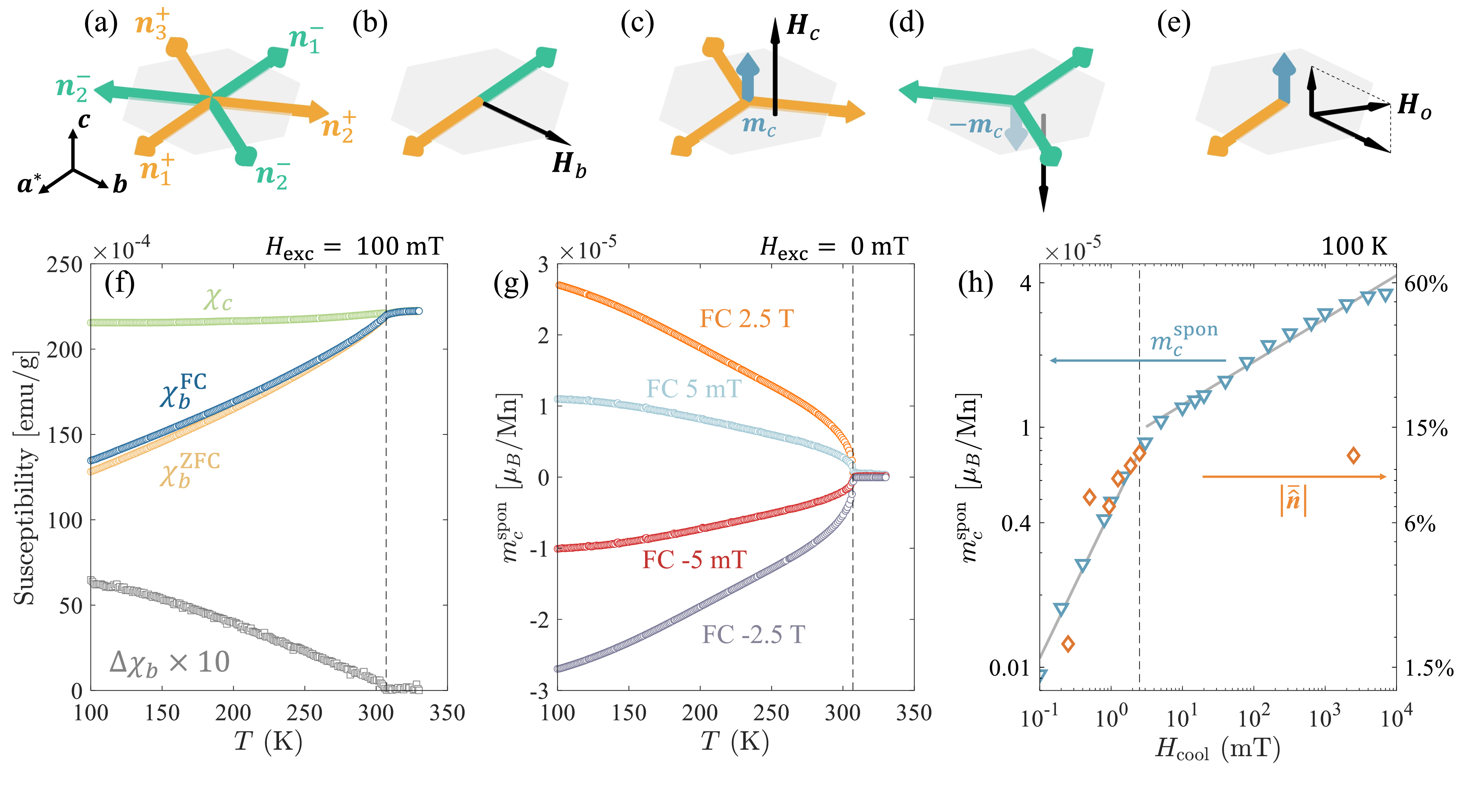}
	\caption{\label{f2} Schematics of (a) a multi-domain state in zero field, (b) a set of orientational domain states selected by an in-plane field $H_b$, (c), (d) Sets of antiphase domain states selected by an out-of-plane field $H_c$, and (e) a single-domain state selected by an oblique field $H_o$.
		(f) Magnetic susceptibilities of MnTe along the $c$-axis ($\chi_c$, green), along the $b$-axis after ZFC ($\chi_b^\text{ZFC}$, yellow), and along the $b$-axis after 5 T FC ($\chi_b^\text{FC}$, blue). The difference $\Delta\chi_b = \chi_b^\text{FC} - \chi_b^\text{ZFC}$ is scaled up by a factor of 10 and shown in gray.
		(g) Spontaneous WFM along the $c$-axis, $m_c^\text{spon}$, in MnTe. The excitation field during the measurement ($H_\text{exc}$) was zero. The data were collected upon warming.
		(h) Cooling-field dependence of $m_c^\text{spon}$ (blue) and $|\overline{\hat{\bm{n}}}|$ (orange) at 100 K, shown on a double-logarithmic scale. The gray solid lines are guides for the eye. The dashed line indicates $H_\text{cool} = 2.5$ mT.}
\end{figure*}

The susceptibility and spontaneous magnetization were measured on bulk MnTe single crystals as a preliminary step toward demonstrating magnetic domain selection by PoND.
Figs.~\ref{f2}(a)-\ref{f2}(e) show schematics of N\'{e}el vectors selected by external magnetic fields. 
As shown in Fig.~\ref{f2}(f), below $T_N = 307$ K, the susceptibility along the $b$-axis after 5-T field cooling (FC), $\chi_b^\text{FC}$, deviates from that after zero-field cooling (ZFC), $\chi_b^\text{ZFC}$, and approaches that along the $c$-axis, $\chi_c$.
This indicates a change from $\chi_\parallel$ to $\chi_\perp$, corresponding to an enhanced population of the single orientational domain state shown in Fig.~\ref{f2}(b).
The spontaneous WFM, $m_c^\text{spon}$, was observed after FC and was found to align with the  cooling field direction, as shown in Fig.~\ref{f2}(g).
The magnitude is around 2$\times$10$^{-5} \mu_\text{B}$ per Mn, consistent with the previous measurements~\cite{kluczykCoexistenceAnomalousHall2024}. 
The ratio of uncompensated DM interactions to the corresponding exchange interaction was estimated from the observed $m_c^\text{spon}$ to be on the order of $10^{-5}$ scale (see Section V in SM~\cite{SM}), supporting interpretation of WFM as a high-order SOC effect as proposed by Ref.~\cite{mazinOriginGossamerFerromagnetism2024}.

The sign of WFM is switchable by the directions of magnetic fields, suggesting the selection of antiphase domain states as shown in Figs.~\ref{f2}(c) and~\ref{f2}(d). 
An oblique field in the $bc$-plane was found to favor a single N\'{e}el vector perpendicular to it, as shown in Fig.~\ref{f2}(e), which was indicated by the magnetic measurements after oblique-field cooling (OFC) as well (see Section II in SM~\cite{SM}). 
The spontaneous WFM is induced by a cooling field at the milli-Tesla scale and tends to saturate at higher fields, as shown in Fig.~\ref{f2}(h).
% On the contrary, although induced by few-Tesla applied field, the spontaneous WFM is not sensitive to the applied field in milli-Tesla scale well below $T_N$, as shown in Fig.~\ref{f2}(i).

\begin{figure*}[htbp]
	\includegraphics[width=\linewidth]{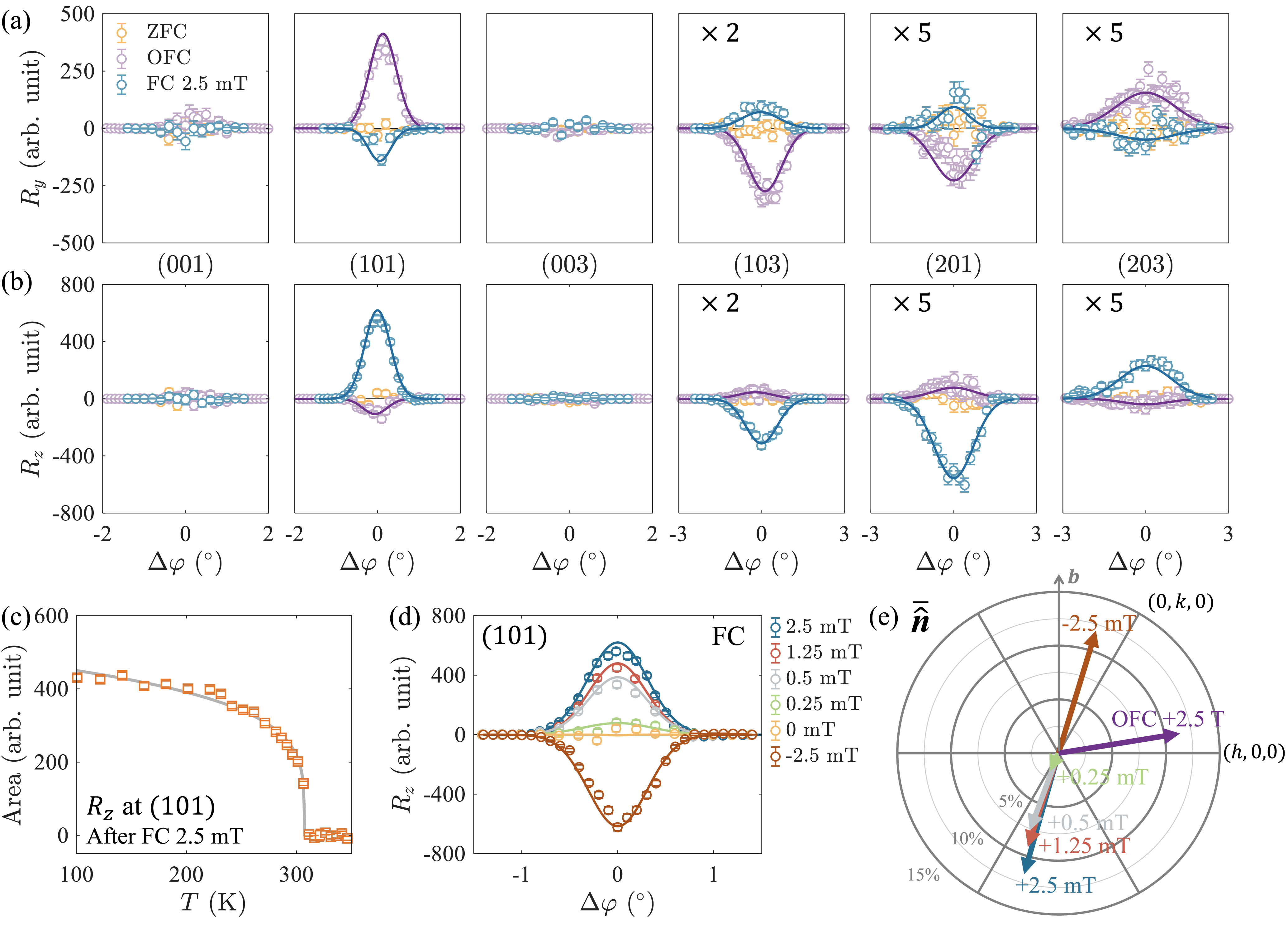}
	\caption{\label{f3}
		Components of the NMI vector, (a) $R_y$ and (b) $R_z$, measured after ZFC, OFC, and FC at the Bragg peaks $(001)$, $(101)$, $(003)$, $(103)$, $(201)$, and $(203)$ at 100 K. Gaussian fits with areas fixed to the calculated values are shown as solid lines. The field in OFC setup was applied in the $bc$-plane as shown in Fig. 2(e), and the field in FC setup was applied in the $c$-axis.
		(c) Temperature dependence of $R_z$ at $(101)$ after 2.5 mT FC. A power-law fit is shown as a solid line.
		(d) Cooling-field dependence of $R_z$ at $(101)$ at 100 K.
		(e) Mapping of $\overline{\hat{\bm{n}}}$ after each cooling condition in the $ab$-plane. 
		The different orientations of $\overline{\hat{\bm{n}}}$ in the FC and OFC setups result in opposite signs of $R_y$ in (a) and $R_z$ in (b).}
\end{figure*}

% \textit{PoND and NMI in {MnTe}}---
Here, referring to the nuclear and magnetic structure factors as $N$ and $\bm{M}$, we define a vector quantity $\bm{R}=(N\bm{M}^\dagger+\bm{M}N^\dagger)$ as a quantitative measure of the NMI effect, whose direction is aligned with $\bm{M}$ and whose modulus reflects the interference amplitude.
In MnTe, the NMI effect happens where the nuclear Bragg peaks coincide with magnetic Bragg peaks, as shown in Figs.~\ref{f1}(c)-\ref{f1}(e). 
Considering the collinear antiferromagnetic spin structure, the NMI vector of MnTe is expressed as
\begin{equation}
	\label{eq1}
	\bm{R}=\bm{\hat{n}}(NM^\dagger+MN^\dagger),
\end{equation}
where $\bm{\hat{n}}$ is the unit N\'{e}el vector (see Section VI in SM~\cite{SM} for details).
In a real MnTe sample with six magnetic domains, the components of $\bm{R}$ are expressed as
\begin{equation}
	\label{eq2}
	R_\alpha=\left[ \left( \sum_{\substack{i=1,2,3\\s=+,-}}p^s_i\bm{\hat{n}}^s_i\right) \cdot\bm{\hat{\alpha}}\right] R,\ \ (\alpha=x,y,z)
\end{equation} 
where $p^s_i$ is the domain population, $\bm{\hat{n}}^s_i$ is the unit N\'{e}el vector of each domain state, and $i$ and $s$ denote the orientational and antiphase domain states, respectively.
Here, a Blume-Maleev coordinate system is adopted, where $\bm{x}$ is parallel to the momentum transfer $\bm{Q}$ and $\bm{y}\perp\bm{x}$ in the scattering plane, as shown in Fig.~\ref{f1}(f).
Therefore, $R_\alpha$ is a direct manifestation of the net unit N\'{e}el vector $\overline{\bm{\hat{n}}}= \sum_{s=+,-}^{i=1,2,3}p^s_i\bm{\hat{n}}^s_i$ in the sample.
 For an ideal single domain, the modulus $\left| \overline{\bm{\hat{n}}}\right| =1$. 
 Note that a balanced distribution of antiphase domains always results in $\left| \overline{\bm{\hat{n}}}\right| =0$, leading to zero $R_\alpha$. 
When the net unit N\'{e}el vector is nonzero, the NMI effect causes a variation in the intensities of scattered neutrons depending on the incident polarization direction, up or down. 
In particular, when the polarization axis of the incident neutrons is parallel to the $y$ ($z$) direction, the difference between the intensities measured with up- and down-incident polarization is proportional to $R_y$ ($R_z$) (see Section III in SM~\cite{SM}).
However, neutrons do not probe the magnetic moment parallel to $\bm{Q}$ (the $\bm{x}$-direction), and therefore $R_x$ cannot be measured.

To probe the altermagnetic order, the PoND experiments were performed on MnTe single crystals in the $(h0l)$ scattering plane using a triple-axis spectrometer. 
Milli-Tesla-scale magnetic fields in arbitrary directions were generated by a group of Helmholtz coils at the sample position, serving as the guide field for polarized neutrons as well as the cooling field for the sample, as illustrated in Fig.~\ref{f1}(f).
The components of the NMI vector, $R_y$ and $R_z$, were unambiguously probed by half-polarized mode at $(101)$, $(103)$, $(201)$, and $(203)$, as shown in Figs.~\ref{f3}(a) and~\ref{f3}(b). Specifically, large $R_y$ and small $R_z$ were observed after OFC (2.5 T along the $c$-axis and 4.3 T along the $b$-axis), small $R_y$ and large $R_z$ after FC along the $c$-axis, and both zero after ZFC.
By contrast, $R_y$ and $R_z$ are always zero at $(001)$ and $(003)$. 
They appear at the superpositions of magnetic and nuclear Bragg peaks with alternating sign, revealing their origin as interference terms, consistent with the calculation in Fig.~\ref{f1}(e).
The temperature dependence of $R_z$ shows critical behavior at $T_N$, as shown in Fig.~\ref{f3}(c), indicating that it contains a magnetic contribution. 
The sign of $R_z$ is switched by FC with fields in the opposite direction, as shown in Fig.~\ref{f3}(d). The magnitude also decreases as the cooling field decreases.

\begin{figure*}[htbp]
	\includegraphics[width=\linewidth]{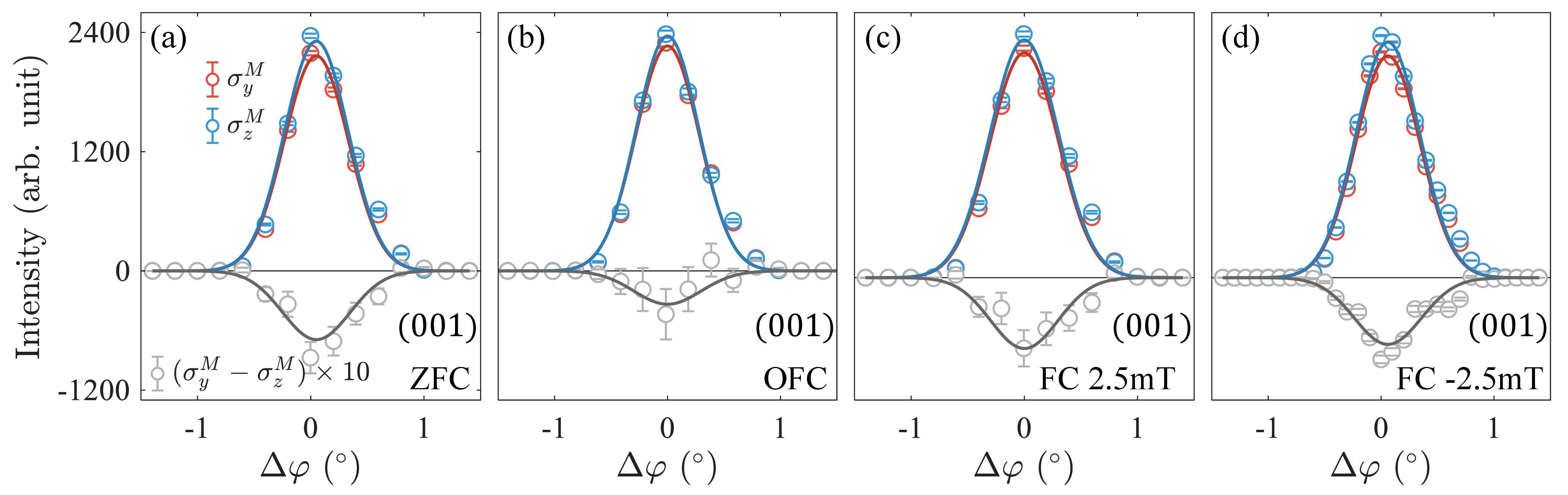}
	\caption{\label{f4}
		$\sigma^M_y$ (red) and $\sigma^M_z$ (blue) obtained using $z$-polarization after (a) ZFC, (b) OFC, (c) FC at 2.5 mT, and (d) FC at $-$2.5 mT at the pure magnetic Bragg peak $(001)$ at 100 K. The difference $\sigma^M_y - \sigma^M_z$ is scaled up by a factor of 10 and shown in gray. Gaussian fits with areas fixed to the calculated values are shown as solid lines.}
\end{figure*} 

The net unit N\'{e}el vector $\overline{\bm{\hat{n}}}$ was obtained from $R_y$ and $R_z$ and mapped in the $ab$-plane (see Section VII in SM~\cite{SM}), as shown in Fig.~\ref{f3}(e).
The $\overline{\bm{\hat{n}}}$ vectors reach the same modulus, approximately 12\%, after OFC at 2.5 T and FC at $\pm2.5$ mT.
They are not strictly aligned with the crystal axes because of the averaging over multiple domains in the sample (see Section VII in SM~\cite{SM}). 
Specifically, after OFC, the $\overline{\bm{\hat{n}}}$ tends to align with the $a^*$-axis, which is perpendicular to the oblique cooling field in the $bc$-plane, indicating an enhanced population of the orientational domain state favored by the field component along the $b$-axis. Despite the relatively small domain imbalance, the $\overline{\bm{\hat{n}}}$ is switched by FC in the opposite direction along the $c$-axis, indicating the field selection of antiphase domain. 
% However, $\overline{\bm{\hat{n}}}$ is induced even without field selection of the orientational domain, suggesting a naturally imbalanced distribution of orientational domains.

The cooling-field dependence of $\left| \overline{\bm{\hat{n}}}\right|$ qualitatively scales with the spontaneous WFM below 2.5 mT, as shown in Fig.~\ref{f2}(h).
In the low-cooling-field region, the magnitude of the spontaneous WFM is attributed to the imbalanced distribution of antiphase domains.
However, in the higher-cooling-field region, the imbalance appears to be saturated, whereas the spontaneous WFM continues to increase. 
In the spontaneous WFM data, a transition point appears around 2.5 mT, and the increase becomes more moderate above this field, which would be related to other mechanism.
% reason: increase of canting angle

The population of one of the orientational domains was further determined from the flipping ratio at a pure magnetic Bragg peak $(001)$ by full-polarized mode with $z$-polarization, as shown in Fig.~\ref{f4}.
In the scattering plane $(h0l)$, an ideally balanced distribution of orientational domains, $p^+_1+p^-_1=p^+_2+p^-_2=p^+_3+p^-_3$, leads to $\sigma^M_y=\sigma^M_z$ (see Section VII in SM~\cite{SM} for details).
Thus, the negative $\sigma^M_y-\sigma^M_z$ observed after both ZFC and FC indicates the presence of a natural imbalance in the domain population, $p^+_1+p^-_1<(p^+_2+p^-_2+p^+_3+p^-_3)/2$, as shown in Figs.~\ref{f4}(a),~\ref{f4}(c), and~\ref{f4}(d), possibly caused by defects and local strain in bulk MnTe.
The natural imbalance was confirmed by a corresponding measurement with $y$-polarization and was found to be common among bulk MnTe samples (see Section IV in SM~\cite{SM}).
$\sigma^M_y$ increases slightly after OFC, as shown in Fig.~\ref{f4}(b), indicating the increase in $p^+_1+p^-_1$, which represents the population of $\bm{n}_1^\pm$ perpendicular to the oblique cooling field, consistent with the spin-flop configuration.

% \textit{Conclusion and Outlook}---
In conclusion, by using MnTe as an example, we show that PoND serves as a powerful probe of altermagnetic order in a bulk sample, complementing the local detection of magnetic antiphase domains by X-ray dichroism~\cite{aminNanoscaleImagingControl2024}, scanning tunneling microscopy~\cite{kaiserMagneticExchangeForce2007}, and optical methods~\cite{schollObservationAntiferromagneticDomains2000,fiebigSecondHarmonicGeneration1994} which are limited to surfaces or thin-film materials.
In addition, the method enables sensitive detection of magnetic order in TRS-breaking antiferromagnets, beyond altermagnets such as MnTe$_2$, thereby promoting research on non-relativistic spin splitting and non-trivial band topology in unconventional antiferromagnets~\cite{yuanPredictionLowZCollinear2021,yuanNonrelativisticSpinSplitting2024,chenUnconventionalMagnonsCollinear2025}.

Our experiments provide direct evidence that the WFM in MnTe is associated with altermagnetic antiphase domains, which can be switched by FC with fields as small as at the milli-Tesla scale. Note that although antiferromagnetic domain walls can give rise to weak ferromagnetism~\cite{bodeAtomicSpinStructure2006a}, this mechanism is unlikely in bulk MnTe: the observed WFM increases as the magnetic domains become more imbalanced, implying a reduced density of domain walls.
WFM is predicted to be widespread in many altermagnets when SOC is included, originating either from spin canting~\cite{autieriStaggeredDzyaloshinskiiMoriyaInteraction2025,roigQuasisymmetryConstrainedSpinFerromagnetism2025} or from orbital ferromagnetism~\cite{joWeakFerromagnetismAltermagnets2025}. 
In addition to MnTe ($2\times10^{-5},\mu_\text{B}$/Mn), spontaneous WFM has been observed to accompany altermagnetic order in FeS ($5\times10^{-4},\mu_\text{B}$/Fe)~\cite{takagiSpontaneousHallEffect2025}, Mn$_5$Si$_3$ ($9\times10^{-3},\mu\text{B}$/Mn)~\cite{baduraObservationAnomalousNernst2025}, and CoF$_2$ ($2\times10^{-3},\mu\text{B}$/Co)~\cite{komuroRevisitingPiezomagneticEffect2025}. 
This phenomenon is akin to the spontaneous magnetization arising from non-collinear or non-coplanar antiferromagnetic orders in Mn$_3$Sn ($2\times10^{-3},\mu\text{B}$/Mn)~\cite{kimataMagneticMagneticInverse2019} and CoTa$_3$S$_6$ ($9\times10^{-4},\mu\text{B}$/Co)~\cite{takagiSpontaneousTopologicalHall2023}, which switch anomalous transport effects. 
Thus, our findings establish a general principle for controlling altermagnetic order and open a pathway toward exploring emergent phenomena in altermagnets, such as GMR-like spin-dependent transport effects~\cite{smejkalGiantTunnelingMagnetoresistance2022,nohTunnelingMagnetoresistanceAltermagnetic2025}. Searching for altermagnets with intrinsic WFM, or inducing WFM through doping or strain, will establish a solid foundation for altermagnet-based spintronics applications.

\begin{acknowledgements}
We are grateful to D. Kawana, T. Asami, and R. Sugiura for their support during the neutron scattering experiments at PONTA.
We also acknowledge R. Ishii for assisting with the sample synthesis. The neutron scattering experiments at PONTA, JRR-3, were performed under user program 2025A. The neutron scattering experiment at HODACA, JRR-3 was performed under user program with proposal No. 25403. Z. Liu was supported by the Japan Society for the Promotion of Science (JSPS) through the Program for Leading Graduate Schools (MERIT). This project was supported by JSPS KAKENHI Grant Numbers 21H04441and 26H02010.
\end{acknowledgements}

{\it Data availability}---The data that support the findings of this article are openly available~\cite{liuObservationofAltermagnetic2026}.

% \bibliography{ref}
%

\end{document}